\documentclass[preprint,authoryear,12pt]{elsarticle}
\usepackage{amssymb}
\usepackage{mathrsfs}        %英文花体字体
\usepackage{epstopdf}% To incorporate .eps illustrations using PDFLaTeX, etc.
\usepackage{amsthm}
\usepackage{caption2}                   %为了将Figure 改成Fig.
\usepackage[colorlinks, citecolor=blue]{hyperref}   %加入蓝色超链接
\hypersetup{colorlinks=false}                       %取消链接的颜色（黑白打印时）
\theoremstyle{plain}% Theorem-like structures
\newtheorem{theorem}{Theorem}

\newtheorem{lemma}{Lemma}

\theoremstyle{definition}
\newtheorem{definition}{Definition}
\newtheorem{example}{Example}
\newtheorem{remark}{Remark} %在中间加上[theorem]会统一编号

\journal{arXiv.org}

\begin{document}

\begin{frontmatter}



\title{A novel class of fixed-time consensus protocols for multi-agent systems with simple dynamics}


\author{Yuquan Chen\corref{cor1}}
\ead{cyq@mail.ustc.edu.cn}%[J. Pan]
%\ead{neudawei@mail.ustc.edu.cn}
\author{Fumian Wang}
%\author{Songsong Cheng}%[Q. Gao]
%\ead{zeng.cb@mail.scut.edu.cn}
%[J. Qiu]
%\author{Yang Yang}%[Q. Gao]
%\ead{jbqiu@hit.edu.cn}
\author{Bing Wang}%[Y. Wang]
\address{Department of Automation, Hohai University, Nanjing, 210024, P.R. China}
%\address[uncc]{Department of Systems Engineering \& Engineering Management, City University of Hong Kong, Tat Chee Avenue, Kowloon, Hong Kong, China}
%\address[scut]{School of Automation Science and Engineering, South China University of Technology, Guangzhou 510640, China}

\cortext[cor1]{Corresponding author}
\phantomsection
\addcontentsline{toc}{section}{Abstract}
\begin{abstract}
  This paper investigates the fixed-time consensus problem for a class of multi-agent systems with simple dynamics. Unlike the traditional way to realize fixed-time convergence, a novel strategy using the property of periodic functions is proposed to achieve fixed-time convergence. On this basis, novel protocols for achieving fixed-time consensus and fixed-time average consensus are then given, where the upper bound of the consensus time is independent of initial conditions. Moreover, the result of fixed-time average consensus is extended to a more general case, where the weights of different states can be allocated in advance. Finally, the fixed-time consensus in the presence of disturbances is derived with the help of sliding mode control, where a fixed-time sliding manifold and fixed-time reaching law are designed. All the conclusions are demonstrated by dedicated simulation examples.
\end{abstract}

\begin{keyword}
 Multi-agent system \sep fixed-time consensus \sep average consensus \sep sliding mode control


\end{keyword}

\end{frontmatter}

% \linenumbers

\section{Introduction}
In the last two decades, multi-agent system has played an important role in all kinds of fields, such as wheeled robots, fixed-wing aircraft, autonomous underwater vehicle, source seeking, smart grid, and so on \citep{ding2017overview,cao2019detecting,valianti2021multi,turner2014distributed,fu2020resilient}. According to different targets, different control expectations are required, such as state consensus, consensus tracking, and formation control. Among them, consensus problem is a basic but important one, where all the agents are expected to have the same behaviour as the time tends to infinity.

Consensus control for multi-agent systems has been studied for quite a long time. A typical linear consensus protocol was proposed by \citet{olfati2004consensus} and it was demonstrated that algebraic connectivity of the interaction graph influences the consensus speed a lot. On this basis, \citet{Kim2006On} provided a proper interaction topology with larger algebraic connectivity to achieve a faster consensus speed. Furthermore, the linear consensus protocol was then extended to a multi-agent system with high-order dynamics \citep{he2011consensus,zhou2014consensus}. Besides, the consensus problem in the presence of disturbances was studied by \citet{mishra2018robust}, where sliding mode control was used to overcome the matched disturbances. A distributed disturbance observer was proposed by \citet{wei2018observer} to estimate the mismatched disturbances and then realized the finite-time tracking. For more details about the consensus problem for multi-agent system, one may refer to the work of \citet{cao2012overview,zuo2018overview} for more details.

In the analyses of consensus problem, faster consensus speed is always desired, for which finite-time consensus problem has gained an increasing attention \citep{wang2010finite,li2011finite,zuo2014new,yu2015finite}. For instance, a finite-time consensus protocol was proposed by \citet{feng2007reaching}, and a finite-time average consensus was also achieved there. On this basis, \citet{wang2010finite} designed a more generalized protocol, where a time-varying order was considered. To further shorten the consensus time, \citep{xiao2009finite} added an additional decaying item based on the work of \citet{feng2007reaching}, and a switching strategy combining existing continuous and discontinuous finite-time consensus protocols was provided by \citet{liu2015finite}. Motivated by the idea of the reaching law design in sliding mode control, \citet{chen2011finite} proposed a discontinuous controller for achieving finite-time consensus, where only sign information of agents' relative states was required.

Although finite-time consensus has been achieved in many ways, the consensus time is explicitly dependent on initial conditions, which is undesirable since the initial conditions are always unknown to us. To solve the problem, fixed-time consensus problem is considered, where there is an upper bound for the consensus time for arbitrary initial conditions. The basic stability theory of fixed-time control was investigated by \citet{polyakov2011nonlinear,polyakov2015finite}. Motivated by the proposed fixed-time stability theory, different fixed-time consensus protocols has been proposed \citep{ning2017distributed,zuo2015nonsingular,fu2016fixed,zuo2017fixed,ning2019practical}. For instance, \citet{zuo2014new} proposed a novel class of fixed-time consensus protocol for multi-agent system with undirected graph, and a fixed-time average consensus was also achieved. On this basis, an improved fixed-time consensus protocol was developed by \citet{zuo2016distributed}, where a less conservative bound for the consensus time was given. The results were then extended to the double-integrator networks by \citet{zuo2015nonsingular}. Recently, the results for fixed-time consensus or tracking has been studied for the multi-agent system with high-order nonlinear dynamics \citep{zuo2017fixed}. Combining with existing fixed-time consensus protocol, fixed-time distributed observer was developed to well solve the tracking problem in the presence of disturbances \citep{wei2018observer,shi2018global}.

Though fixed-time consensus problem has been studied for such a long time and many complex situations have been considered, the main idea for achieving fixed-time consensus is quite similar, which is motivated by the idea of \citet{polyakov2011nonlinear,polyakov2015finite}. Different from the fixed-time theory proposed by \citet{polyakov2011nonlinear,polyakov2015finite}, we will propose a new class of fixed-time theory, where the property of sine function is used to achieve fixed-time convergence. On this basis, novel protocols for achieving fixed-time consensus and fixed-time average consensus for a class of multi-agent systems with simple dynamics are then given. All the results are finally extended to the multi-agent system with disturbances with the help of sliding mode control. Contributions of this paper are concluded as follows:
\begin{itemize}
  \item A novel strategy using the property of periodic function is proposed to realize fixed-time convergence.
  \item A novel protocol is given to achieve fixed-time consensus for a class of multi-agent system with simple dynamics.
  \item A novel protocol for achieving fixed-time average consensus is provided, and then extended to a more general case.
  \item Fixed-time consensus protocol for multi-agent with disturbances is designed by using the sliding mode control, where a fixed-time sliding manifold and reaching law are given.
\end{itemize}

\textbf{Notations:} Throughout this paper, we use a weighted undirected graph $\mathscr G(A)=\{\mathscr V,E,A\}$ to describe the information exchanging among the agents, where $\mathscr V(\mathscr{G})=\{v_1,v_2,\cdots,v_n\}$ is the node set, $A=[a_{ij}]_{n\times n}$ is the adjacent matrix, where $a_{ij}=a_{ji}$ for undirected graph, and $E(\mathscr G)$ is the edge set. The set of all neighbours of node $v_i$ is denoted by ${N_i} = \left\{ {{v_j}:\left( {{v_i},{v_j}} \right) \in E} \right\}$. $f(t)*g(t)$ denotes the convolution of function $f(t)$ and $g(t)$.
\section{Problem Formulation}
Consider a group of $n$ agents with integrator like dynamics in the form of
\begin{eqnarray}\label{sys1}
\dot x_i(t)=u_i(t),~i=1,2,\cdots,n,
\end{eqnarray}
where, $x_i\in {\mathbb R}^N$ and $u_i\in {\mathbb R}^N$ denote the state and control input called protocol of agent $i$, respectively. In this study, $N=1$ is assumed for simplicity in the sequel. The results to be developed later are valid for the vector case by introducing Kronecker product.

\begin{definition}\citep{zuo2014new}
System (\ref{sys1}) is said to achieve fixed-time consensus if for any initial state $x_i(0)$ and any $i,j\in {\mathscr V}$, there exist a protocol and a globally bounded function $T_0\ge 0$ such that
\begin{eqnarray}
\left\{ \begin{array}{l}
{\lim _{t \to {T_0}}}\left| {{x_i}\left( t \right) - {x_j}\left( t \right)} \right| = 0,\\
\left| {{x_i}\left( t \right) - {x_j}\left( t \right)} \right| = 0,\forall t > {T_0}.
\end{array} \right.
\end{eqnarray}
Moreover, if the final consensus state satisfies ${x_i}\left( t \right) = \frac{1}{n}\sum\nolimits_{i = 1}^n {{x_i}\left( 0 \right)} $ for all $i\in {\mathscr V}$ and $t>T_0$, it is said to achieve the fixed-time average consensus.
\end{definition}

\begin{lemma}
Following inequality always holds
\[{\left( {\sum\limits_{i = 1}^n {{x_i}} } \right)^2} \le n\sum\limits_{i = 1}^n {x_i^2} ,\]
where $x_i\in {\mathbb R},~i=1,2,\cdots,n$.
\end{lemma}
\section{A novel strategy for fixed-time convergence}
In this section, unlike the traditional way to derive fixed-time convergence \citep{basin2016continuous}, we propose a novel algorithm that guarantees fixed-time convergence. Consider the following algorithm
\begin{eqnarray}\label{ftpro}
\left\{ \begin{array}{l}
\dot x =  - \theta {\mathop{\rm sgn}} \left( x \right),\\
\dot \theta  =  - \lambda \theta  + \rho \left| x \right|,
\end{array} \right.
\end{eqnarray}
where $\theta(0)=0$, $x,~\theta\in{\mathbb R}$, $\lambda>0$ and $\rho>0$ are tuning parameters.
\begin{theorem}\label{them1}
Algorithm (\ref{ftpro}) achieves fixed-time convergence if $\rho>\frac{\lambda^2}{4}$. Moreover, the convergence time satisfies
\[t_r\le \frac{\pi }{{\sqrt {\rho  - \frac{{{\lambda ^2}}}{4}} }}.\]
\end{theorem}
\begin{proof}
Consider the Lyapunov function $V=|x|$. Take time derivative of $V$, yielding, $\dot V =  - \theta.$
Then one has
\begin{eqnarray}\label{eq1}
\left\{ \begin{array}{l}
\dot V =  - \theta, \\
\dot \theta  =  - \lambda \theta  + \rho V.
\end{array} \right.\end{eqnarray}
Perform Laplace transform of (\ref{eq1}) on both sides, arriving at
\begin{eqnarray}\label{15}
\left\{ \begin{array}{l}
sV\left( s \right) - V\left( 0 \right) =  - \theta \left( s \right),\\
s\theta \left( s \right) =  - \lambda \theta \left( s \right) + \rho V\left( s \right).
\end{array} \right.
\end{eqnarray}

One can then get $V(s)$ by solving (\ref{15})
\[V\left( s \right) = \frac{{s + \lambda }}{{s\left( {s + \lambda } \right) + \rho }}V\left( 0 \right).\]
Perform inverse Laplace transform of $V(s)$ and define ${\omega : = \sqrt {\rho  - \frac{{{\lambda ^2}}}{4}} }$, yielding,
\begin{eqnarray}
V\left( t \right) = {e^{ - \frac{\lambda }{2}t}}\left( {\cos \left( {\omega t} \right) + \frac{\lambda }{\omega }\sin \left( {\omega t} \right)} \right)V\left( 0 \right),
\end{eqnarray}
 from which it is concluded that $V(t)$ must reach zero in a half cycle, i.e., $\frac{\pi}{\omega}$. Furthermore, since $\theta(0)=0$ and $V(t)\ge 0$, it is found that $\theta(t)\ge 0$ for all $t\ge 0$ and $\dot V(t)\le 0$ then follows. Finally, combining with that $V(t)\ge 0$ and $\dot V(t)\le 0$, it can be proven that $V(t)$ will stay as zero thereafter once $V(t)$ reaches zero, which implies a fixed-time convergence. This completes the proof.
\end{proof}
\begin{remark}
Unlike the traditional way to derive the fixed-time convergence, the property of sine function, whose first positive zero is determined by the frequency (irrelevant to initial conditions), is used to guarantee a fixed-time convergence. Moreover, the additional item $-\lambda \theta$ is used to attenuating the amplitude of $\theta$ after fixed-time convergence, which can help attenuating the chattering phenomenon for practical usage, which will be shown in the simulation.
\end{remark}
Then consider the following more general case
\begin{eqnarray}\label{eq2}
\left\{ \begin{array}{l}
\dot x \le  - \theta, \\
\dot \theta  \ge  - \lambda \theta  + \rho x,
\end{array} \right.
\end{eqnarray}
where $\theta(0)=0$, $x(t)\ge 0$ for all $t$, $\lambda>0$ and $\rho>0$ are tuning parameters.
\begin{theorem}\label{them2}
Algorithm (\ref{eq2}) achieves fixed-time convergence if $\rho>\frac{\lambda^2}{4}$. Moreover, the convergence time satisfies
\[t_r\le \frac{\pi }{{\sqrt {\rho  - \frac{{{\lambda ^2}}}{4}} }}.\]
\end{theorem}
\begin{proof}
One can transform (\ref{eq2}) into the following form
\begin{eqnarray}
\left\{ \begin{array}{l}
\dot x =  - \theta  - m\left( t \right),\\
\dot \theta  =  - \lambda \theta  + \rho x + n\left( t \right),
\end{array} \right.
\end{eqnarray}
where $m(t)\ge 0$ and $n(t)\ge 0$ for all $t\ge 0$.

Performing Laplace transform on both sides, yields,
\begin{eqnarray}
\left\{ \begin{array}{l}
sx\left( s \right) - x\left( 0 \right) =  - \theta \left( s \right) - m\left( s \right),\\
s\theta \left( s \right) =  - \lambda \theta \left( s \right) + \rho x\left( s \right) + n\left( s \right).
\end{array} \right.
\end{eqnarray}

One can then get $x(s)$ as
\[x\left( s \right) = \frac{{s + \lambda }}{{{s^2} + \lambda s + \rho }}\left( {x\left( 0 \right) - m\left( s \right)} \right) - \frac{{n\left( s \right)}}{{{s^2} + \lambda s + \rho }},\]
and furthermore the solution $x(t)$ in time domain can then be derived
\begin{eqnarray}
x\left( t \right) = {e^{ - \frac{\lambda }{2}t}}\left( {\cos \left( {\omega t} \right) + \frac{\lambda }{\omega }\sin \left( {\omega t} \right)} \right)*\left( {x\left( 0 \right) - m\left( t \right)} \right) - \frac{{{e^{ - \frac{\lambda }{2}t}}}}{\omega }\sin \left( {\omega t} \right)*n\left( t \right),
\end{eqnarray}
where ${\omega : = \sqrt {\rho  - \frac{{{\lambda ^2}}}{4}} }$.

Suppose $t_0$ is the first positive zero of function ${\cos \left( {\omega t} \right) + \frac{\lambda }{\omega }\sin \left( {\omega t} \right)}$, which is smaller than $\frac{\pi}{\omega}$. We have $\sin(\omega t)\ge 0$ and ${\cos \left( {\omega t} \right) + \frac{\lambda }{\omega }\sin \left( {\omega t} \right)}\ge 0$ for all $0<t\le t_0$. Combining with the fact that $m(t)\ge 0$ and $n(t)\ge 0$, it is concluded that
\begin{eqnarray}
x\left( t \right){ \le ^{ - \frac{\lambda }{2}t}}\left( {\cos \left( {\omega t} \right) + \frac{\lambda }{\omega }\sin \left( {\omega t} \right)} \right)x\left( 0 \right),
\end{eqnarray}
which implies that $x(t)$ reaches zero within $t_r\le t_0\le \frac{\pi}{\omega}$.

Combining with the fact $\theta(t)\ge 0$ and thus $\dot x(t)\le 0$, it is concluded that $x(t)$ will stay on zero thereafter once it reaches zero. Moreover, the reaching time is smaller than $\frac{\pi}{\omega}$ for arbitrary initial conditions. This completes the proof.
\end{proof}
\section{Fixed-time consensus protocol}
In this section, a novel fixed-time consensus protocol is provided based on Theorem \ref{them1} and Theorem \ref{them2}, with which fixed-time consensus for system (\ref{sys1}) can be derived. The protocol can be formulated as
\begin{eqnarray}\label{pro1}
\left\{ \begin{array}{l}
{u_i} = {\theta _i}{\left( {\sum\nolimits_{j \in {N_i}} {{a_{ij}}\left( {{x_j} - {x_i}} \right)} } \right)^{ - 1}},\\
{\dot \theta _i} =  - \lambda {\theta _i} + \rho {\sum\nolimits_{j \in {N_i}} {{a_{ij}}\left( {{x_j} - {x_i}} \right)} ^2},
\end{array} \right.
\end{eqnarray}
where $\theta_i(0)=0$, $\lambda>0$, $\rho>0$ are tuning parameters.
\begin{theorem}\label{them3}
If the undirected graph $\mathscr G(A)$ of system (\ref{sys1}) is connected, then protocol (\ref{pro1}) achieves the fixed-time consensus if $\rho>\frac{\lambda^2}{16}$. Moreover, the consensus time satisfies
\[{t_r} \le \frac{\pi }{{\sqrt {4\rho  - \frac{{{\lambda ^2}}}{4}} }}.\]
\end{theorem}
\begin{proof}
Consider the following semi-positive definite function
\begin{eqnarray}
V = \frac{1}{4}{\sum\limits_{i = 1}^n {\sum\limits_{j = 1}^n {{a_{ij}}\left( {{x_j} - {x_i}} \right)}^2 }}.
\end{eqnarray}
Here, since ${\mathscr G}(A)$ is connected, $V=0$ can imply the consensus of all states (\citep{zuo2014new}). Take the time derivative of $V$, and one has
\begin{eqnarray}
\begin{array}{rl}
\dot V =& \sum\limits_{i = 1}^n {\frac{{\partial V}}{{\partial {x_i}}}{{\dot x}_i}} \\
 =& \sum\limits_{i = 1}^n {{\theta _i}\left( { - \sum\limits_{j = 1}^n {{a_{ij}}\left( {{x_j} - {x_i}} \right)} } \right){{\left( {\sum\nolimits_{j \in {N_i}} {{a_{ij}}\left( {{x_j} - {x_i}} \right)} } \right)}^{ - 1}}} \\
 =&  - \sum\limits_{i = 1}^n {{\theta _i}}.
\end{array}
\end{eqnarray}

Besides, one can obtain following equality
\begin{eqnarray}
\begin{array}{rl}
\sum\limits_{i = 1}^n {{{\dot \theta }_i}}  =&  - \lambda \sum\limits_{i = 1}^n {{\theta _i}}  + \rho \sum\limits_{i = 1}^n {\sum\nolimits_{j \in {N_i}} {{a_{ij}}{{\left( {{x_j} - {x_i}} \right)}^2}} } \\
 =&  - \lambda \sum\limits_{i = 1}^n {{\theta _i}}  + 4\rho V.
\end{array}
\end{eqnarray}
Define $\Theta : = \sum\limits_{i = 1}^n {{\theta _i}} $, following equalities can be obtained
\begin{eqnarray}
\left\{ \begin{array}{l}
\dot V =  - \Theta, \\
\dot \Theta  =  - \lambda \Theta  + 4\rho V.
\end{array} \right.
\end{eqnarray}

According to Theorem \ref{them1}, it is concluded that fixed-time convergence of $V(t)$ can be guaranteed if $\rho>\frac{\lambda^2}{16}$, which indicates the fixed-time consensus. And the convergence time satisfies
\[{t_r} \le \frac{\pi }{{\sqrt {4\rho  - \frac{{{\lambda ^2}}}{4}} }}.\]
This completes the proof.
\end{proof}
\begin{remark}
Some comments on Theorem \ref{them3} are given as follows.
\begin{itemize}
  \item Fixed-time consensus can be guaranteed according to Theorem \ref{them1}, where the upper bound of the consensus time is dependent on the designed parameters in protocol (\ref{pro1}). However, it cannot guarantee an average consensus since $\sum\limits_{i = 1}^n {{{\dot x}_i}\left( t \right)}  \ne 0$.
  \item To avoid singularity, one can replace ${\left( {\sum\nolimits_{j \in {N_i}} {{a_{ij}}\left( {{x_j} - {x_i}} \right)} } \right)^{ - 1}}$ with
      \begin{eqnarray}\label{12}
      \frac{{\sum\nolimits_{j \in {N_i}} {{a_{ij}}\left( {{x_j} - {x_i}} \right)} }}{{{{\left( {\sum\nolimits_{j \in {N_i}} {{a_{ij}}\left( {{x_j} - {x_i}} \right)} } \right)}^2} + \gamma }},~\gamma>0,
      \end{eqnarray}
      which is similar to that using a saturation function to replace sign function in sliding mode control.
  \item To achieve fixed-time consensus for arbitrary initial conditions, input saturation must be taken into consideration. For practical usage, one can enlarge the value of $\gamma$ in (\ref{12}) to attenuate the amplitude of the control input.
\end{itemize}
\end{remark}

\section{Fixed-time average consensus protocol}
In this section, an average consensus protocol is designed as
\begin{eqnarray}\label{pro2}
\left\{ \begin{array}{l}
{u_i} = \sum\nolimits_{j \in {N_i}} {{a_{ij}}\left( {{\theta _i} + {\theta _j}} \right){{\left( {{x_j} - {x_i}} \right)}^{ - 1}}}, \\
{\dot \theta _i} =  - \lambda {\theta _i} + \rho \sum\nolimits_{j \in {N_i}} {{a_{ij}}\left( {{x_j} - {x_i}} \right)^2},
\end{array} \right.
\end{eqnarray}
where $\theta_i(0)=0$, $\lambda>0$ and $\rho>0$ are tuning parameters.
\begin{theorem}\label{them4}
If the undirected graph $\mathscr G(A)$ of system (\ref{sys1}) is connected, then protocol (\ref{pro2}) achieves the fixed-time average consensus if $\rho>\frac{n}{8\kappa^2}\lambda^2$. Moreover, the consensus time satisfies
\[{t_r} \le \frac{\pi }{{\sqrt {\frac{2\kappa^2\rho}{n}  - \frac{{{\lambda ^2}}}{4}} }},\]
where $\kappa  = \mathop {\min }\limits_{{a_{ij}} \ne 0} {a_{ij}}$.
\end{theorem}
\begin{proof}
According to the designed protocol (\ref{pro2}), it is found that
\[\sum\limits_{i = 1}^n {{{\dot x}_i}}  = \sum\limits_{i = 1}^n {\sum\nolimits_{j \in {N_i}} {{a_{ij}}\left( {{\theta _i} + {\theta _j}} \right){{\left( {{x_j} - {x_i}} \right)}^{ - 1}}} }  = 0\]
because of the symmetry. Then define ${x^*} = \frac{1}{n}\sum\limits_{i = 1}^n {{x_i}\left( t \right)}$, which is time invariant, and let $\delta_i(t)=x_i(t)-x^*(t)$. Consider the following Lyapunov function
\begin{eqnarray}\label{lya}
V = \frac{1}{2}\sum\limits_{i = 1}^n {\delta _i^2}.
\end{eqnarray}
Differentiating $V$ versus time, yields,
\begin{eqnarray}
\begin{array}{rl}
\dot V =& \sum\limits_{i = 1}^n {{\delta _i}{{\dot \delta }_i}} \\
 =& \sum\limits_{i = 1}^n {{\delta _i}\sum\nolimits_{j \in {N_i}} {{a_{ij}}\left( {{\theta _i} + {\theta _j}} \right){{\left( {{x_j} - {x_i}} \right)}^{ - 1}}} } \\
 =& \frac{1}{2}\sum\limits_{i = 1}^n {\sum\limits_{j = 1}^n {{a_{ij}}\left( {{\theta _i} + {\theta _j}} \right)\left( {{\delta _i} - {\delta _j}} \right){{\left( {{\delta _j} - {\delta _i}} \right)}^{ - 1}}} } \\
 =& -\frac{1}{2}\sum\limits_{i = 1}^n {\sum\limits_{j = 1}^n {{a_{ij}}\left( {{\theta _i} + {\theta _j}} \right)} } \\
 \le&  - \kappa \sum\limits_{i = 1}^n {{\theta _i}}.
\end{array}
\end{eqnarray}
where $\kappa  = \mathop {\min }\limits_{i,j,{a_{ij}} \ne 0} {a_{ij}}$. On the other hand, we have
\begin{eqnarray}
V = \frac{1}{2}\sum\limits_{i = 1}^n {\delta _i^2}  = \frac{1}{2}\sum\limits_{i = 1}^n {{{\left( {\sum\limits_{j = 1}^n {\frac{{{x_i} - {x_j}}}{n}} } \right)}^2}}  \le \frac{1}{{2n}}\sum\limits_{i = 1}^n {\sum\limits_{j = 1}^n {{{\left( {{x_i} - {x_j}} \right)}^2}} }.
\end{eqnarray}

Since the graph ${\mathscr G}(A)$ is connected, it is concluded that for arbitrary $k~{\rm and}~r=1,2,\cdots,n,$
\begin{eqnarray}
\kappa\left( {{x_k} - {x_r}} \right)^2 \le \sum\limits_{i = 1}^n {\sum\nolimits_{j \in {N_i}} {{a_{ij}}{{\left( {{x_j} - {x_i}} \right)}^2}} }.
\end{eqnarray}
Therefore, the following inequality can be derived
\[\begin{array}{rl}
\sum\limits_{i = 1}^n {\sum\nolimits_{j \in {N_i}} {{a_{ij}}{{\left( {{x_j} - {x_i}} \right)}^2}} }  =& \frac{1}{{{n^2}}}{n^2}\sum\limits_{i = 1}^n {\sum\nolimits_{j \in {N_i}} {{a_{ij}}{{\left( {{x_j} - {x_i}} \right)}^2}} } \\
 \ge& \frac{\kappa}{{{n^2}}}\sum\limits_{i = 1}^n {\sum\limits_{j = 1}^n {{{\left( {{x_i} - {x_j}} \right)}^2}} } \\
 \ge& \frac{\kappa}{n}\sum\limits_{i = 1}^n {{{\left( {\sum\limits_{j = 1}^n {\frac{{{x_i} - {x_j}}}{n}} } \right)}^2}} \\
 =& \frac{2\kappa}{{n}}V.
\end{array}\]

Define $\Theta : = \kappa\sum\limits_{i = 1}^n {{\theta _i}} $ and we arrive at
\begin{eqnarray}
\left\{ \begin{array}{l}
\dot V \le  - \Theta, \\
\dot \Theta  \ge  - \lambda \Theta  + \frac{2\kappa^2\rho }{n}V.
\end{array} \right.
\end{eqnarray}

According to the results of Theorem \ref{them2}, it is concluded that fixed-time convergence of $V(t)$ can be guaranteed if $\rho>\frac{n}{8\kappa^2}\lambda^2$, which indicates the fixed-time consensus. And the convergence time satisfies
\[{t_r} \le \frac{\pi }{{\sqrt {\frac{2\kappa^2\rho}{n}  - \frac{{{\lambda ^2}}}{4}} }}.\]
This completes the proof.
\end{proof}
\begin{remark}
Some comments on Theorem \ref{them4} are given as follows.
\begin{itemize}
  \item The gain $\theta_i+\theta_j$ instead of $\theta_i$ is used in protocol (\ref{pro2}) to guarantee the symmetry property, and then we have that $\sum\limits_{i = 1}^n {{{\dot x}_i}\left( t \right)}  = 0$, which helps achieving fixed-time average consensus.
  \item For practical usage, following protocol without singularity can be applied
  \begin{eqnarray}\label{14}
  {u_i} = \sum\nolimits_{j \in {N_i}} {{a_{ij}}\left( {{\theta _i} + {\theta _j}} \right)\frac{{{x_j} - {x_i}}}{{{{\left( {{x_j} - {x_i}} \right)}^2} + \gamma }}},~\gamma>0,
  \end{eqnarray}
  where $\gamma$ can be designed properly to decrease the amplitude of control input.
  \item The condition $\rho>\frac{n}{8\kappa^2}\lambda^2$ is only a sufficient condition. Therefore, some $\rho>0$ does not satisfy such condition may still guarantee a fixed-time average consensus, which will be validated in the simulation.
\end{itemize}
\end{remark}
The result of fixed-time average consensus in Theorem \ref{them4} can be extended to a more general case. Consider the following consensus protocol
\begin{eqnarray}\label{pro3}
\left\{ \begin{array}{l}
{u_i} = \sum\nolimits_{j \in {N_i}} {\frac{{{a_{ij}}}}{{{p_i}}}\left( {{\theta _i} + {\theta _j}} \right){{\left( {{x_j} - {x_i}} \right)}^{ - 1}}}, \\
{\theta _i} =  - \lambda {\theta _i} + \rho \sum\nolimits_{j \in {N_i}} {{a_{ij}}{{\left( {{x_j} - {x_i}} \right)}^2}},
\end{array} \right.
\end{eqnarray}
where $\theta_i(0)=0$, $p_i>0$, $\sum\limits_{i = 1}^n {{p_i} = 1} $, $\lambda>0$ and $\rho>0$ are tuning parameters.
\begin{theorem}\label{them5}
If the undirected graph $\mathscr G(A)$ of system (\ref{sys1}) is connected, then protocol (\ref{pro3}) achieves the fixed-time consensus to ${x^*} = \sum\limits_{i = 1}^n {{p_i}{x_i}\left( 0 \right)} $ if $\rho>\frac{K^3n^3}{8\kappa^2}\lambda^2$. Moreover, the consensus time satisfies
\[{t_r} \le \frac{\pi }{{\sqrt {\frac{2\kappa^2\rho}{K^3n^3}  - \frac{{{\lambda ^2}}}{4}} }},\]
where $\kappa  = \mathop {\min }\limits_{{a_{ij}} \ne 0} {a_{ij}}$ and $K = \mathop {\max }\limits_{i} {p_{i}}$.
\end{theorem}
\begin{proof}
According to the designed protocol (\ref{pro3}), it is found that
\[\sum\limits_{i = 1}^n {{{p_i\dot x}_i}}  = \sum\limits_{i = 1}^n {\sum\nolimits_{j \in {N_i}} {{a_{ij}}\left( {{\theta _i} + {\theta _j}} \right){{\left( {{x_j} - {x_i}} \right)}^{ - 1}}} }  = 0.\]
Let ${x^*} = \sum\limits_{i = 1}^n {{p_i}{x_i}\left( t \right)} $ which is time invariant and define $\delta_i=x_i(t)-x^*$. Consider the same Lyapunov function as (\ref{lya}). Differentiating $V$ versus time, yields,
\begin{eqnarray}
\begin{array}{rl}
\dot V =& \sum\limits_{i = 1}^n {{\delta _i}{{\dot \delta }_i}} \\
 =& \sum\limits_{i = 1}^n {{\delta _i}\sum\nolimits_{j \in {N_i}} {\frac{a_{ij}}{p_i}\left( {{\theta _i} + {\theta _j}} \right){{\left( {{x_j} - {x_i}} \right)}^{ - 1}}} } \\
 =& \frac{1}{2}\sum\limits_{i = 1}^n {\sum\limits_{j = 1}^n {\frac{a_{ij}}{p_i}\left( {{\theta _i} + {\theta _j}} \right)\left( {{\delta _i} - {\delta _j}} \right){{\left( {{\delta _j} - {\delta _i}} \right)}^{ - 1}}} } \\
 =& -\frac{1}{2}\sum\limits_{i = 1}^n {\sum\limits_{j = 1}^n {\frac{a_{ij}}{p_i}\left( {{\theta _i} + {\theta _j}} \right)} } \\
 \le&  - \frac{\kappa}{K} \sum\limits_{i = 1}^n {{\theta _i}}.
\end{array}
\end{eqnarray}
On the other hand, one has
\[\begin{array}{rl}
\sum\limits_{i = 1}^n {\sum\nolimits_{j \in {N_i}} {{a_{ij}}{{\left( {{x_j} - {x_i}} \right)}^2}} }  =& \frac{1}{{{n^2}}}{n^2}\sum\limits_{i = 1}^n {\sum\nolimits_{j \in {N_i}} {{a_{ij}}{{\left( {{x_j} - {x_i}} \right)}^2}} } \\
 \ge& \frac{\kappa }{{{n^2}}}\sum\limits_{i = 1}^n {\sum\limits_{j = 1}^n {{{\left( {{x_i} - {x_j}} \right)}^2}} } \\
 \ge& \frac{\kappa }{{K^2{n^2}}}\sum\limits_{i = 1}^n {\sum\limits_{j = 1}^n {p_i^2{{\left( {{x_i} - {x_j}} \right)}^2}} } \\
 \ge& \frac{\kappa }{{K^2{n^3}}}\sum\limits_{i = 1}^n {{{\left( {\sum\limits_{j = 1}^n {{p_i}\left( {{x_i} - {x_j}} \right)} } \right)}^2}} \\
 =& \frac{{2\kappa }}{{K^2{n^3}}}V.
\end{array}.\]
Define $\Theta : = \frac{\kappa}{K}\sum\limits_{i = 1}^n {{\theta _i}} $ and we arrive at \begin{eqnarray}
\left\{ \begin{array}{l}
\dot V \le  - \Theta, \\
\dot \Theta  \ge  - \lambda \Theta  + \frac{{2{\kappa ^2}\rho }}{{{K^3}{n^3}}}V.
\end{array} \right.
\end{eqnarray}
According to the results of Theorem \ref{them2}, it is concluded that fixed-time convergence of $V(t)$ can be guaranteed if $\rho>\frac{K^3n^3}{8\kappa^2}\lambda^2$, which indicates the fixed-time consensus. And the convergence time satisfies
\[{t_r} \le \frac{\pi }{{\sqrt {\frac{2\kappa^2\rho}{K^3n^3}  - \frac{{{\lambda ^2}}}{4}} }}.\]
This completes the proof.
\end{proof}
\begin{remark}
Some comments on Theorem \ref{them5} are given as follows
\begin{itemize}
  \item Theorem \ref{them5} will reduce to Theorem \ref{them4} when $p_i=\frac{1}{n}$. The upper bound of the consensus time is then $\frac{\pi }{{\sqrt {{2\kappa^2\rho} - \frac{{{\lambda ^2}}}{4}} }}$, which is different from that of Theorem \ref{them4}. The reason is that an extra gain $\frac{1}{p_i}$ is included in protocol (\ref{pro3}).
  \item The condition in $\rho>\frac{K^3n^3}{8\kappa^2}\lambda^2$ is still a sufficient one. The strategy (\ref{14}) for eliminating the singularity for protocol (\ref{pro2}) can also be applied.
\end{itemize}
\end{remark}
\section{Fixed-time consensus in the presence of disturbance}
In this section, system (\ref{sys1}) with disturbances is considered, which can be formulated as
\begin{eqnarray}\label{sys2}
{\dot x_i} = {u_i} + {\Delta _i},
\end{eqnarray}
where $\Delta_i,~i=1,2,\cdots,n$ are bounded disturbances. In the following, it is assumed that $d = \mathop {\max }\limits_i \left| {{\Delta _i}} \right|$ is known in prior. To eliminate the influence of the disturbances, sliding mode control is used, where a novel fixed-time sliding manifold and reaching law are proposed. The consensus protocol can be formulated as
\begin{eqnarray}\label{pro4}
\left\{ \begin{array}{l}
{u_i} = \sum\nolimits_{j \in {N_i}} {{{{a_{ij}}}}\left( {{\theta _i} + {\theta _j}} \right){{\left( {{x_j} - {x_i}} \right)}^{ - 1}} - \left( {{\eta _i} + d} \right){\mathop{\rm sgn}} \left( {{s_i}} \right)}, \\
{\theta _i} =  - \lambda {\theta _i} + \rho \sum\nolimits_{j \in {N_i}} {{a_{ij}}{{\left( {{x_j} - {x_i}} \right)}^2}}, \\
{s_i} = {x_i} - \int_0^t {\sum\nolimits_{j \in {N_i}} {{{{a_{ij}}}}\left( {{\theta _i} + {\theta _j}} \right){{\left( {{x_j} - {x_i}} \right)}^{ - 1}}d\tau } }, \\
{{\dot \eta }_i} =  - \omega {\eta _i} + \mu \left| {{s_i}} \right|,
\end{array} \right.
\end{eqnarray}
where $\theta_i(0)=0,~\eta_i(0)=0$, and $\lambda>0,~\omega>0,~\rho>0,~\mu>0$ are tuning parameters.
\begin{theorem}\label{them6}
If the undirected graph $\mathscr G(A)$ of system (\ref{sys2}) is connected, then protocol (\ref{pro4}) achieves the fixed-time consensus to ${x^*} = 0$ if $\mu> \frac{\omega^2}{4}$ and $\rho>\frac{n}{8\kappa^2}\lambda^2$. Moreover, the consensus time satisfies
\[{t_r} \le \frac{\pi }{{\sqrt {\mu  - \frac{{{\omega ^2}}}{4}} }}+\frac{\pi }{{\sqrt {\frac{2\kappa^2\rho}{n}  - \frac{{{\lambda ^2}}}{4}} }},\]
where $\kappa  = \mathop {\min }\limits_{{a_{ij}} \ne 0} {a_{ij}}$
\end{theorem}
\begin{proof}
The proof can be divided into three steps. Step one is to prove that the sliding manifold can be reached in a fixed time. Consider the following Lyapunov function
\[V = \sum\limits_{i = 1}^n {\left| {{s_i}} \right|}. \]
Differentiating $V$ versus time, yields,
\begin{eqnarray}
\begin{array}{rl}
\dot V =& \sum\limits_{i = 1}^n {{{\dot s}_i}{\mathop{\rm sgn}} \left( {{s_i}} \right)} \\
 =& \sum\limits_{i = 1}^n {\left( {{{\dot x}_i} - \sum\nolimits_{j \in {N_i}} {\frac{{{a_{ij}}}}{{{p_i}}}\left( {{\theta _i} + {\theta _j}} \right){{\left( {{x_j} - {x_i}} \right)}^{ - 1}}} } \right){\mathop{\rm sgn}} \left( {{s_i}} \right)} \\
 =& \sum\limits_{i = 1}^n {\left( { - \left( {{\eta _i} + d} \right){\mathop{\rm sgn}} \left( {{s_i}} \right) + {\Delta _i}} \right){\mathop{\rm sgn}} \left( {{s_i}} \right)} \\
 =& \sum\limits_{i = 1}^n {\left( { - {\eta _i} + d - {\Delta _i}{\mathop{\rm sgn}} \left( {{s_i}} \right)} \right)} \\
 \le&  - \sum\limits_{i = 1}^n {{\eta _i}}.
\end{array}
\end{eqnarray}

Besides, define $\Xi  = \sum\limits_{i = 1}^n {{\eta _i}} $  and we have that
\begin{eqnarray}
\begin{array}{rl}
\dot \Xi  =& \sum\limits_{i = 1}^n {{{\dot \eta }_i}} \\
 =& \sum\limits_{i = 1}^n {\left( { - \omega {\eta _i} + \mu \left| {{s_i}} \right|} \right)} \\
 =&  - \omega \sum\limits_{i = 1}^n {{\eta _i}}  + \mu \sum\limits_{i = 1}^n {\left| {{s_i}} \right|} \\
 =&  - \omega \Xi  + \mu V.
\end{array}
\end{eqnarray}

Then following inequalities can be obtained
\begin{eqnarray}
\left\{ \begin{array}{l}
\dot V \le  - \Xi, \\
\dot \Xi  =  - \omega \Xi  + \mu V.
\end{array} \right.
\end{eqnarray}

According to Theorem \ref{them2}, if $\mu>\frac{\omega^2}{4}$, fixed-time convergence of $V$ can be derived and the reaching time satisfies
\[{t_s} \le \frac{\pi }{{\sqrt {\mu  - \frac{{{\omega ^2}}}{4}} }},\]
which implies the fixed-time convergence to the sliding manifold.

Step two is to prove the fixed-time consensus after reaching the sliding manifold. After reaching the sliding manifold, one has that
\[\begin{array}{l}
{{\dot s}_i} = {{\dot x}_i} - \sum\nolimits_{j \in {N_i}} {{{{a_{ij}}}}\left( {{\theta _i} + {\theta _j}} \right){{\left( {{x_j} - {x_i}} \right)}^{ - 1}}}  = 0\\
 \Rightarrow {{\dot x}_i} = \sum\nolimits_{j \in {N_i}} {{{{a_{ij}}}}\left( {{\theta _i} + {\theta _j}} \right){{\left( {{x_j} - {x_i}} \right)}^{ - 1}}}.
\end{array}\]
The closed-loop system of (\ref{sys2}) can then switched to protocol (\ref{pro3}). According to Theorem \ref{them4}, fixed-time consensus is guaranteed in less than $\frac{\pi }{{\sqrt {\frac{2\kappa^2\rho}{n}  - \frac{{{\lambda ^2}}}{4}} }}$. However, the condition $\sum\limits_{i = 1}^n {{{\dot x}_i}\left( t \right)}  = 0$ cannot be guaranteed due to the existing disturbances, and thus the consensus state is not $\sum\limits_{i = 1}^n {{{\dot x}_i}\left( 0 \right)}$ any more.

The last step is to find the consensus state $x^*$. After reaching the sliding manifold, we have that $s_i(t)=\dot s_i(t)=0$, which indicates
\begin{eqnarray}
\left\{ \begin{array}{l}
{x_i} = \int_0^t {\sum\nolimits_{j \in {N_i}} {{a_{ij}}\left( {{\theta _i} + {\theta _j}} \right){{\left( {{x_j} - {x_i}} \right)}^{ - 1}}d\tau } }, \\
{{\dot x}_i} = \sum\nolimits_{j \in {N_i}} {{a_{ij}}\left( {{\theta _i} + {\theta _j}} \right){{\left( {{x_j} - {x_i}} \right)}^{ - 1}}}.
\end{array} \right.
\end{eqnarray}

On this basis, it is concluded that
\begin{eqnarray}\label{con1}
\sum\limits_{i = 1}^n {{x_i}\left( t \right)}  = 0~{\rm{and}}~\sum\limits_{i = 1}^n {{{\dot x}_i}\left( t \right)}  = 0.
\end{eqnarray}
Combining with the fixed-time consensus of all states, the proposed protocol can guarantee a fixed-time consensus to $x^*=0$. Moreover, the consensus time is less than
\[{t_r} \le \frac{\pi }{{\sqrt {\mu  - \frac{{{\omega ^2}}}{4}} }}+\frac{\pi }{{\sqrt {\frac{2\kappa^2\rho}{n}  - \frac{{{\lambda ^2}}}{4}} }}.\]
This completes the proof.
\end{proof}
\begin{remark}
Protocol (\ref{pro4}) can guarantee the fixed-time consensus to the origin. Furthermore, by modifying the sliding manifold as
\begin{eqnarray}\label{11}
{s_i} = {x_i} - \bar x_i - \int_0^t {\sum\nolimits_{j \in {N_i}} {{a_{ij}}\left( {{\theta _i} + {\theta _j}} \right){{\left( {{x_j} - {x_i}} \right)}^{ - 1}}d\tau } },
\end{eqnarray}
where $\bar x_i$ are designed constant parameters, the consensus state in Theorem \ref{them6} will then be
\[{x^*} = \frac{1}{n}\sum\limits_{i = 1}^n {\bar x_i}. \]
The conclusion can be directly derived since the following conditions (similar to (\ref{con1}) hold after reaching the sliding manifold
\[\sum\limits_{i = 1}^n {{x_i}\left( t \right)}  = \sum\limits_{i = 1}^n {\bar x_i} ~{\rm{and}}~\sum\limits_{i = 1}^n {{{\dot x}_i}\left( t \right)}  = 0.\]
According to the fixed-time consensus of all states, the conclusion is achieved.
\end{remark}
\section{Illustrative examples}
In this section, simulation examples are provided to demonstrate all the proposed conclusions. Consider a six-agent system in a network with a connection matrix
\[\left[ {\begin{array}{*{20}{c}}
{\begin{array}{*{20}{c}}
0&1\\
1&0
\end{array}}&{\begin{array}{*{20}{c}}
0&0\\
1&0
\end{array}}&{\begin{array}{*{20}{c}}
1&1\\
1&0
\end{array}}\\
{\begin{array}{*{20}{c}}
0&1\\
0&0
\end{array}}&{\begin{array}{*{20}{c}}
0&1\\
1&0
\end{array}}&{\begin{array}{*{20}{c}}
0&0\\
0&0
\end{array}}\\
{\begin{array}{*{20}{c}}
1&1\\
1&0
\end{array}}&{\begin{array}{*{20}{c}}
0&0\\
0&0
\end{array}}&{\begin{array}{*{20}{c}}
0&0\\
0&0
\end{array}}
\end{array}} \right].\]
%. In the simulations, we set $a_{ij}=1$ for simplicity.
%\begin{figure}
%\centering
%\includegraphics[width=0.8\textwidth]{ff1.pdf}
%\caption{Undirected communication topology of system (\ref{sys1})}\label{f7}
%\end{figure}
\begin{example}\label{ex1}
In this example, we will show the fixed-time consensus with protocol (\ref{pro1}). Set $\rho=2$ and $\lambda=2$. Consider different initial conditions as
\[\left\{ \begin{array}{l}
{\rm {case}}~1:~x\left( 0 \right) = {\left[ { - 5, 2,4,-2,-4,5} \right]^{\rm T}},\\
{\rm {case}}~2:~x\left( 0 \right) = {\left[ {10, - 20, - 3,9,4, - 30} \right]^{\rm T}}.
\end{array} \right.\]
Results are shown in Fig \ref{f1} and Fig \ref{f2}. Fixed-time consensus can be observed in both cases. Moreover, the estimated upper bound for the consensus time is $\frac{\pi}{\sqrt{4\rho-\frac{\lambda^2}{4}}}\approx 1.19$ (sec). The consensus time for two different initial conditions are very close (about 0.83 (sec)), which is smaller than the estimated consensus time and implies the fixed-time consensus. However, the consensus state is dependent on initial conditions, which cannot guarantee an average consensus.
\begin{figure}
\centering
\includegraphics[width=0.8\textwidth]{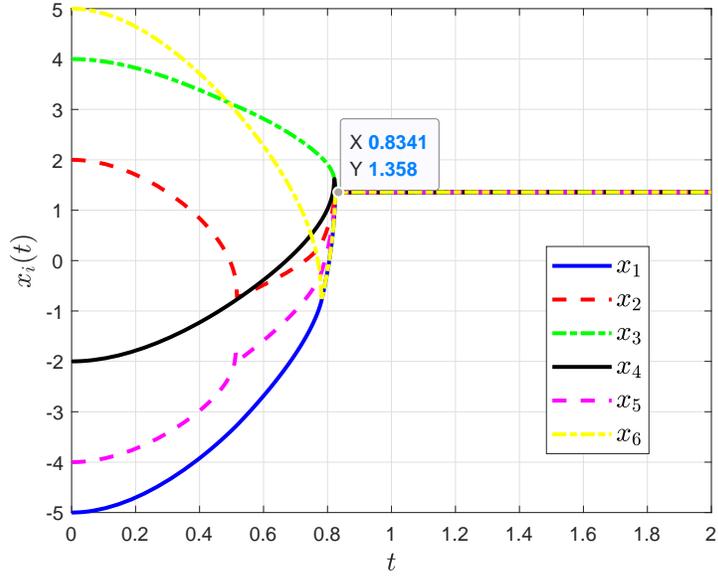}
\caption{Fixed-time consensus with case 1 in Example \ref{ex1}}\label{f1}
\end{figure}
\begin{figure}
\centering
\includegraphics[width=0.8\textwidth]{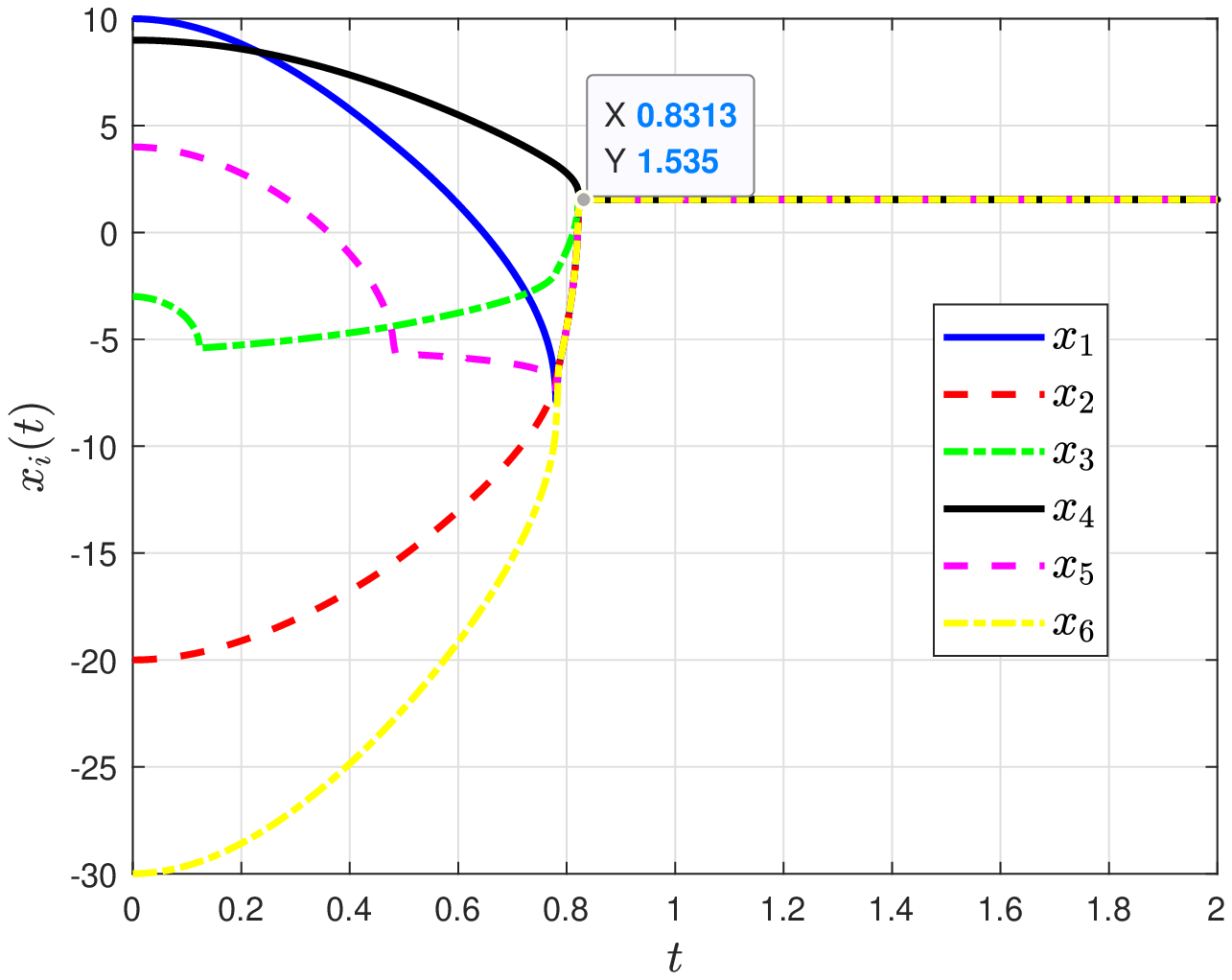}
\caption{Fixed-time consensus with case 2 in Example \ref{ex1}}\label{f2}
\end{figure}
\end{example}

\begin{example}\label{ex2}
In this example, we will show the fixed-time average consensus with protocol (\ref{pro2}). Set $\rho=8$ and $\lambda=2$. Consider the same initial conditions as Example \ref{ex1}. Results are shown in Fig \ref{f3} and Fig \ref{f4}. Fixed-time consensus can be observed in both cases. Moreover, the estimated upper bound for the consensus time is $\frac{\pi}{\sqrt{\frac{2\rho}{n}-\frac{\lambda^2}{4}}}\approx 2.43$ (sec). The consensus time for two different initial conditions (0.18 (sec) and 0.31 (sec)) are smaller than the estimated consensus time. Moreover, the average consensus is achieved for both cases, where the average states are $0$ and $-5$ respectively. Since the condition for fixed-time consensus is sufficient, it is found that protocol with $\rho=2$ (does not satisfy condition $\rho>\frac{n}{8\kappa^2}\lambda^2$) can still guarantee a fixed-time average consensus as shown in Fig \ref{f5}.
\begin{figure}
\centering
\includegraphics[width=0.8\textwidth]{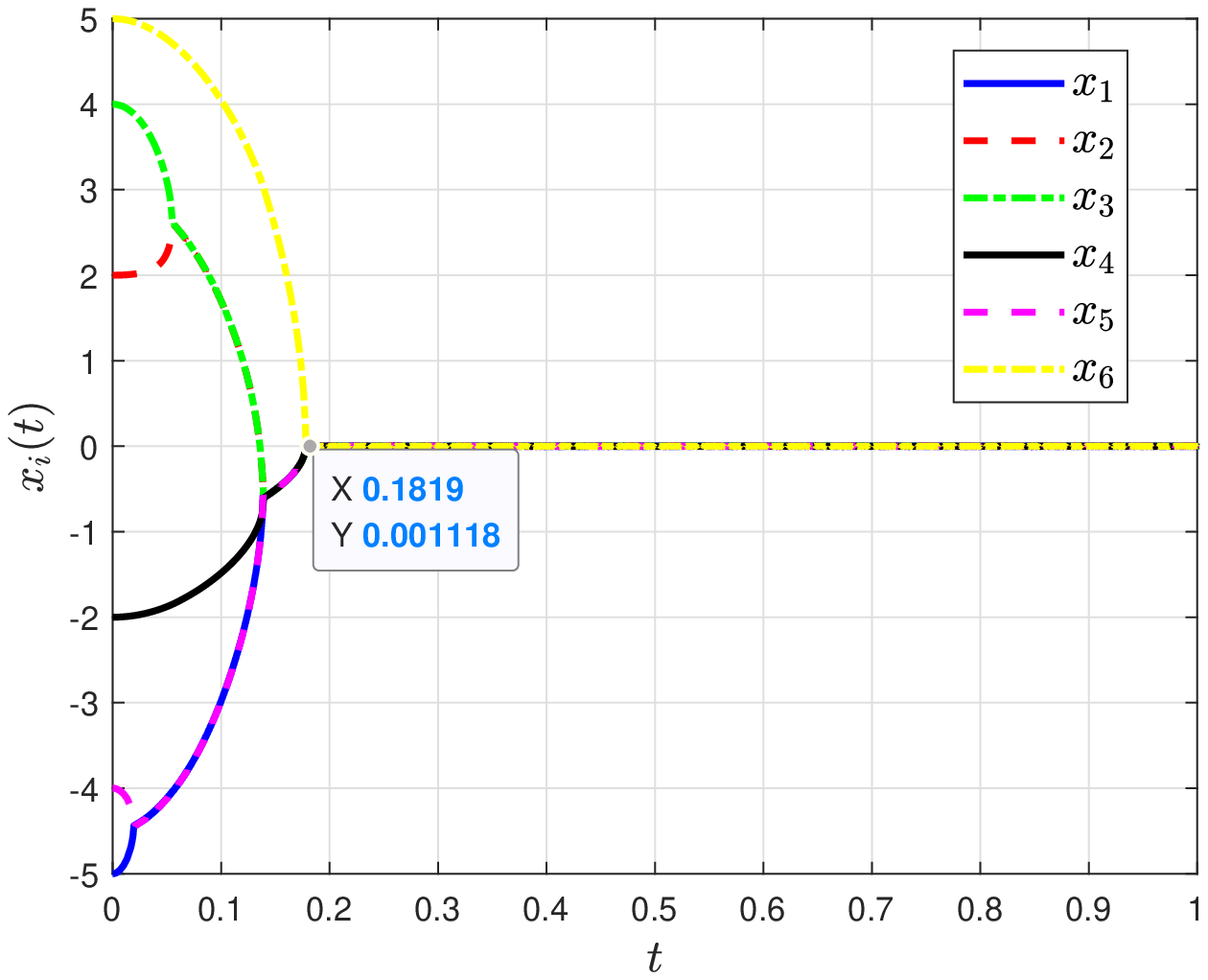}
\caption{Fixed-time consensus with case 1 and $\rho=8$ in Example \ref{ex2}}\label{f3}
\end{figure}
\begin{figure}
\centering
\includegraphics[width=0.8\textwidth]{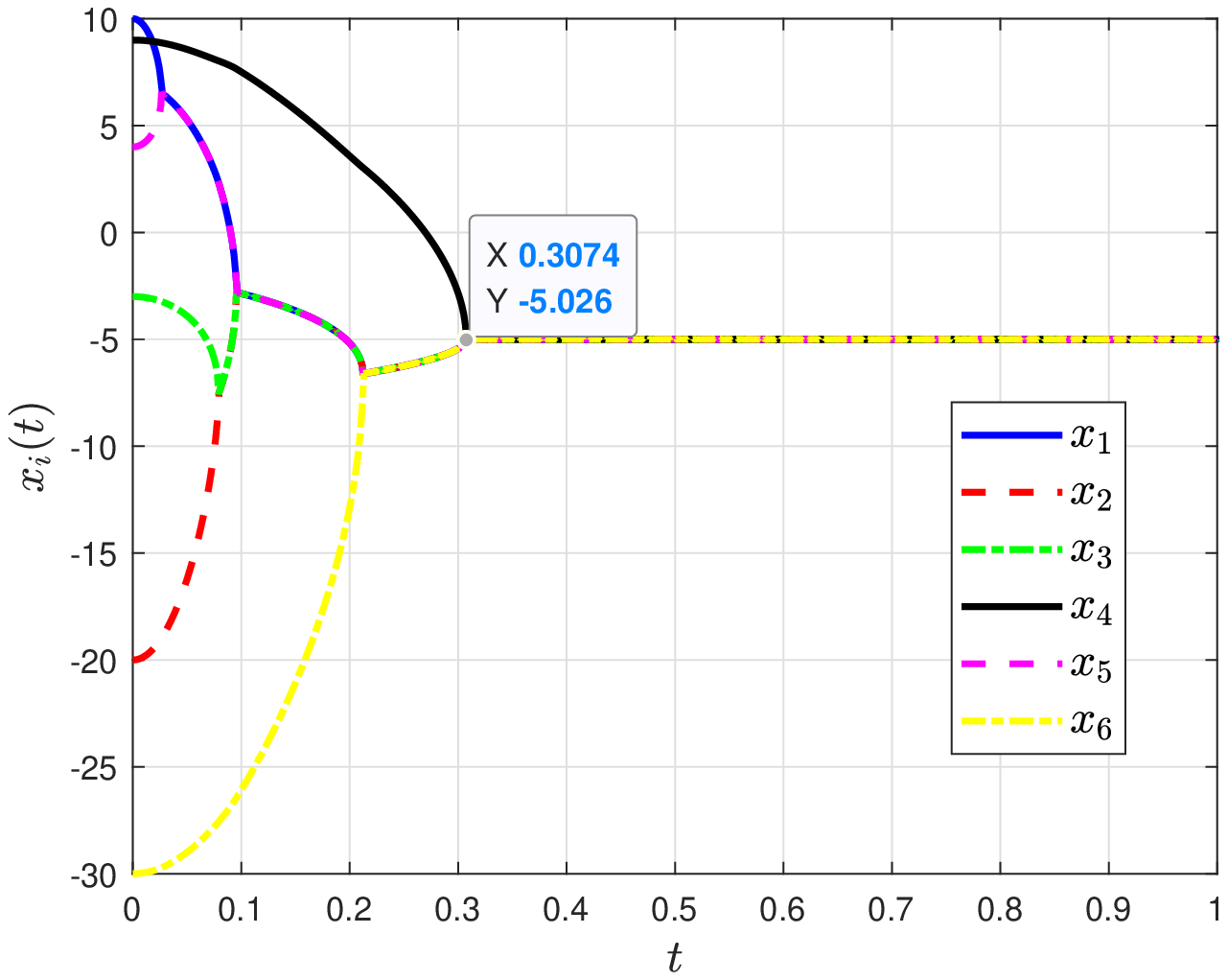}
\caption{Fixed-time consensus with case 2 and $\rho=8$ in Example \ref{ex2}}\label{f4}
\end{figure}

\begin{figure}
\centering
\includegraphics[width=0.8\textwidth]{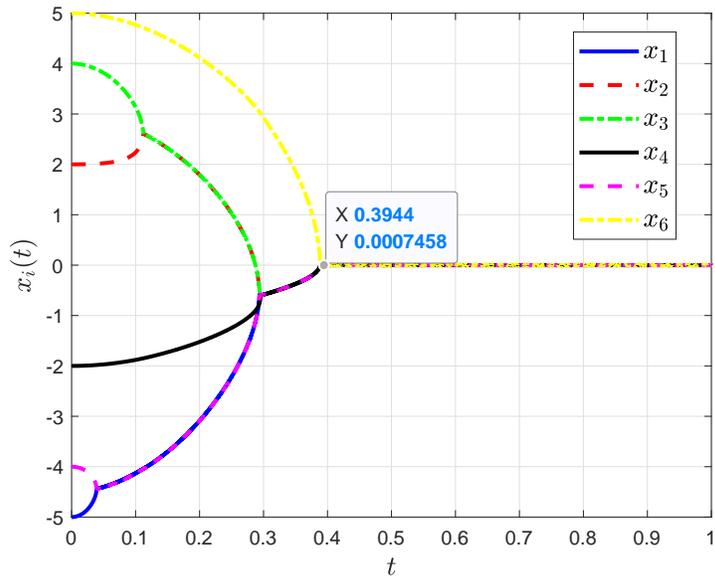}
\caption{Fixed-time consensus with case 1 and $\rho=2$ in Example \ref{ex2}}\label{f5}
\end{figure}
%\begin{figure}
%\centering
%\includegraphics[width=0.8\textwidth]{ex2-f4.eps}
%\caption{Fixed-time consensus with case 2 in Example \ref{ex2}}\label{f6}
%\end{figure}
\end{example}
\begin{example}\label{ex3}
In this example, protocol (\ref{pro3}) is considered where $p_1=p_2=\frac{1}{12}$, $p_3=p_4=\frac{1}{6}$, $p_5=p_6=\frac{1}{4}$. Initial condition is set as $x\left( 0 \right) = {\left[ {12, - 12,6,6,4,4} \right]^{\rm T}}$, where $x^*=2$. When simulating, set $\lambda=2$ and $\rho=1$. Results are shown in Fig \ref{f6}. It is observed that fixed-time consensus to $x^*=2$ can be guaranteed. We have to declare that $\rho=1$ does not satisfy the condition in Theorem \ref{them5}, but it can still guarantee the fixed-time consensus since the condition is only a sufficient one.
\begin{figure}
\centering
\includegraphics[width=0.8\textwidth]{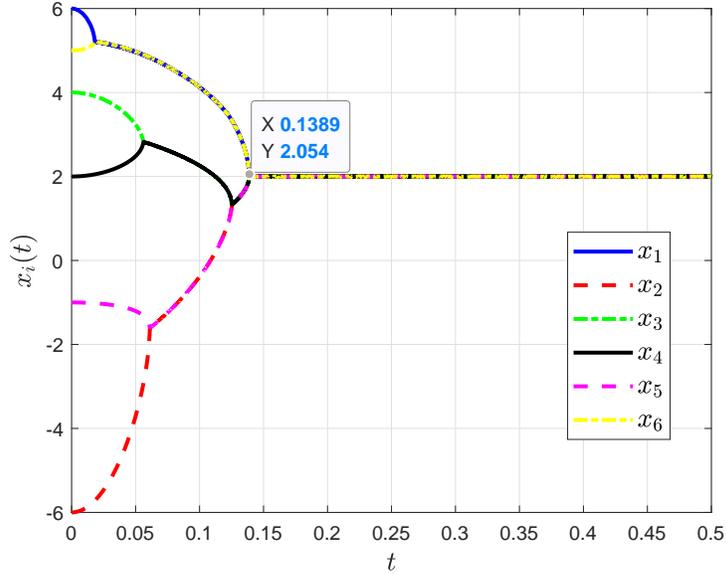}
\caption{Fixed-time consensus with case 1 in Example \ref{ex3}}\label{f6}
\end{figure}
\end{example}

\begin{example}\label{ex4}
In this example, protocol (\ref{pro4}) is considered. Initial condition is set as $x\left( 0 \right) = {\left[ {1, -2,3,-4,5,-6} \right]^{\rm T}}$, and disturbances are set as
\[\Delta  = {\left[ {\sin \left( {10t} \right),0.8\sin \left( {10t} \right),0.5\sin \left( {10t} \right),\cos \left( {10t} \right),0.8\cos \left( {10t} \right),0.5\cos \left( {10t} \right)} \right]^{\rm T}}.\]
When simulating, set $\lambda=2$, $\rho=0.4$, $\omega=4$, $\mu=10$, and $d=1$ (the upper bound of the disturbances). Results are shown in Fig \ref{f7}-Fig\ref{f9}. Following observations can be directly derived.
\begin{itemize}
  \item Fixed-time consensus to the origin is achieved according to Fig \ref{f7} and the consensus time is about $0.87$ (sec).
  \item Fixed-time convergence to the sliding manifold is achieved according to Fig \ref{f8}. Moreover, the reaching time is about $0.81$ (sec), which is shorter than the estimated one, i.e., $\frac{\pi}{\sqrt{\mu-\frac{\omega^2}{4}}}\approx 1.28$ (sec).
  \item To avoid singularity, $\gamma=0.01$ in (\ref{14}) is used. The maximum amplitude of control input is acceptable (about $42.5$) according to Fig \ref{f9}. Moreover, the fixed-time convergence to the sliding manifold does not worsen the chattering phenomenon a lot since the amplitude of the chattering is attenuated to $d=1$ quickly.
\end{itemize}
\begin{figure}
\centering
\includegraphics[width=0.8\textwidth]{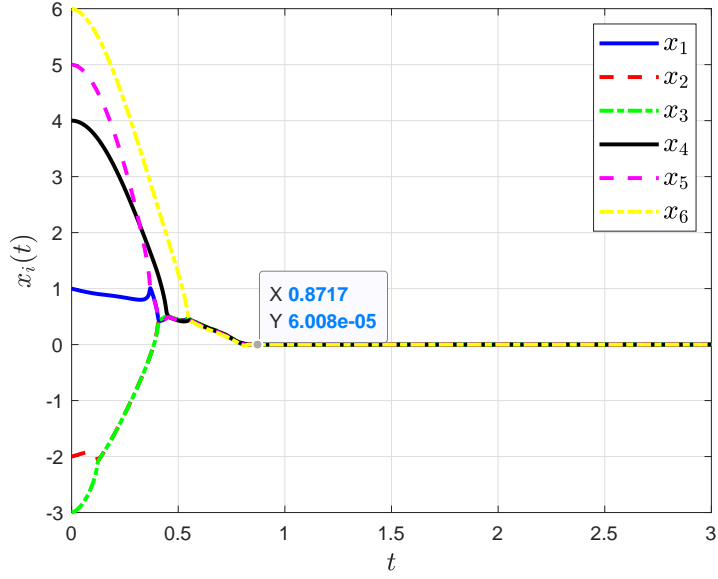}
\caption{Fixed-time consensus in the presence of disturbances in Example \ref{ex4}}\label{f7}
\end{figure}
\begin{figure}
\centering
\includegraphics[width=0.8\textwidth]{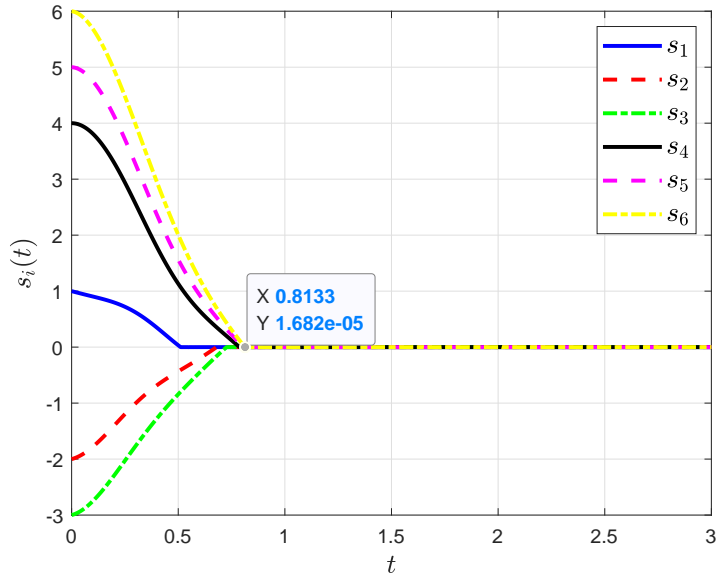}
\caption{Fixed-time convergence to sliding manifold in Example \ref{ex4}}\label{f8}
\end{figure}
\begin{figure}
\centering
\includegraphics[width=0.8\textwidth]{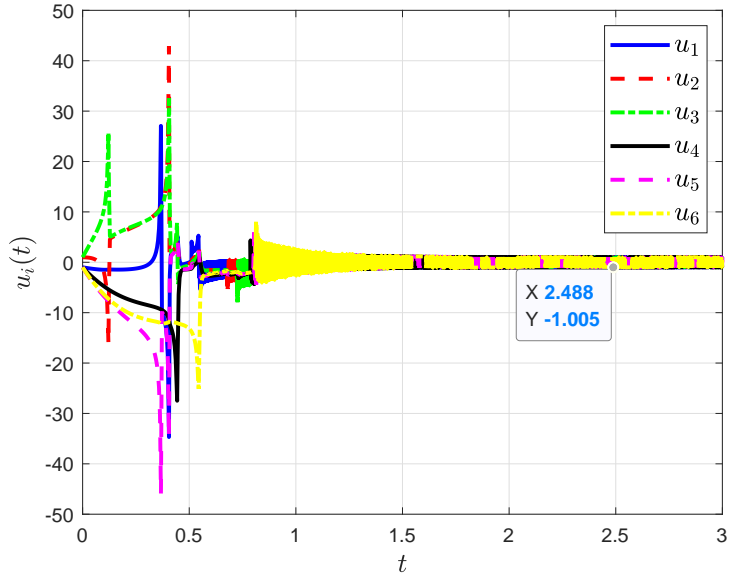}
\caption{Control inputs in Example \ref{ex3}}\label{f9}
\end{figure}
\begin{figure}
\centering
\includegraphics[width=0.8\textwidth]{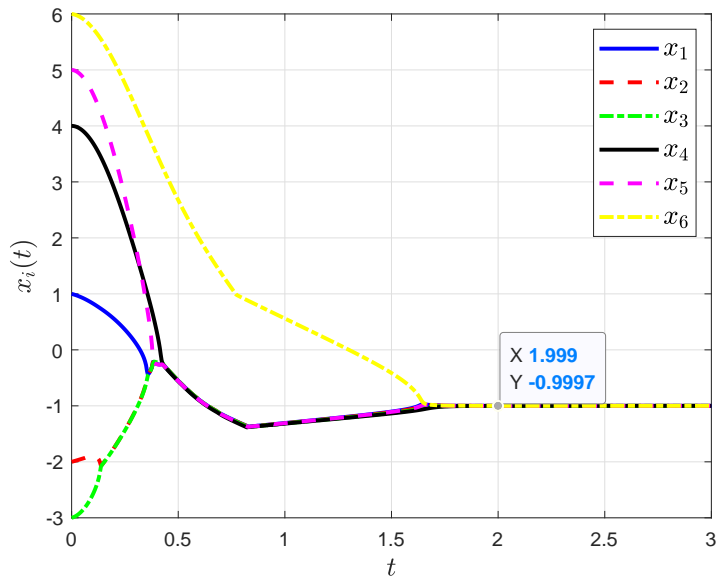}
\caption{Fixed-time consensus with sliding manifold (\ref{11}) in Example \ref{ex4}}\label{f10}
\end{figure}

By using the sliding manifold \ref{11} with $\bar x = {\left[ {-2,0,-2,-2,-2,2} \right]^{\rm T}}$, results are shown in Fig \ref{f10}, where fixed-time consensus to $x^*=-1$ is achieved.
\end{example}
\section{Conclusions}
In this paper, fixed-time consensus problem for a class of multi-agent systems has been studied. A novel strategy to realize fixed-time convergence is provided at first. Then, novel protocols for achieving fixed-time consensus and average consensus are then given. Finally, the fixed-time consensus in the presence of disturbances is obtained with the help of sliding mode control, where both a fixed-time sliding manifold and reaching law are designed. All the conclusions are validated by simulation examples. There are some promising directions for future research:
\begin{itemize}
  \item extend the results to the multi-agent system with high-order dynamics;
  \item design novel fixed-time distributed disturbance observer to overcome the mismatched disturbances.
\end{itemize}
\phantomsection
\addcontentsline{toc}{section}{References}
\bibliographystyle{model5-names}
\bibliography{fixedtimeconsensus}

%% not want to use model5-names.bst.



%% \bibitem must have one of the following forms:
%%   \bibitem[Jones et al.(1990)]{key}...
%%   \bibitem[Jones et al.(1990)Jones, Baker, and Williams]{key}...
%%   \bibitem[Jones et al., 1990]{key}...
%%   \bibitem[\protect\citeauthoryear{Jones, Baker, and Williams}{Jones
%%       et al.}{1990}]{key}...
%%   \bibitem[\protect\citeauthoryear{Jones et al.}{1990}]{key}...
%%   \bibitem[\protect\astroncite{Jones et al.}{1990}]{key}...
%%   \bibitem[\protect\citename{Jones et al., }1990]{key}...
%%   \harvarditem[Jones et al.]{Jones, Baker, and Williams}{1990}{key}...
%%


% \end{thebibliography}

\end{document}